\documentclass[12pt]{iopart}
\usepackage{amsfonts}
\begin{document}
\jl{1}
\title{ Noncommutative $3D$ harmonic oscillator}

\author{Anais Smailagic\dag{}\footnote[1]{e-mail
        address: \texttt{anais@ictp.trieste.it}},
        Euro Spallucci\ddag{}\footnote[5]{e-mail
        address: \texttt{spallucci@trieste.infn.it}}
        }
        \address{\dag{}Sezione INFN di Trieste,\\
         Strada Costiera 11, 34014 Trieste,\\
         Italy}
        \address{\ddag{}Department of Theoretical Physics,\\
         University of Trieste, Strada Costiera 11, 34014 Trieste,\\
         Italy}
		        
\date{\today}

\begin{abstract}
We find transformation matrices allowing to express
non-commutative three dimensional harmonic oscillator in terms
of an isotropic commutative oscillator, following 
``philosophy of simplicity'' approach. Non-commutative parameters
have  physical interpretation in terms  of an external magnetic field. 
Furthermore, we show that for a particular choice of  noncommutative parameters
there is an equivalent anisotropic representation, whose transformation matrices
are far more complicated. We indicate a way to obtain the more complex solutions
from the simple ones.
\end{abstract}

\pacno{03.65G, 03.65F, 02.40G, 11.15-q}
\maketitle
 String theory results \cite{witten}, \cite{sw}, have generated a revival
 of interest for field theory in a non-commutative geometry \cite{aref}. 
 A simpler insight
 on the role of non-commutativity in field theory can be obtained studying
 solvable models of non-commutative quantum mechanics \cite{jell},
 \cite{bellucci},\cite{baner}.\\
 Recently, we have presented \cite{noi} the  description of the
 noncommutative harmonic oscillator in two dimensions in terms of an isotropic
 commutative oscillator in an external magnetic field. This interpretation is
 made possible by the existence of a simple representation of the 
 noncommutative coordinates in terms of the canonical ones. There are many other
 possible representations of the noncommutative algebra in terms of two
 Heisenberg algebras \cite{nair}. Nevertheless, all of them fall in two groups: 
 those leading to a set of anisotropic oscillators, and others leading to an 
 isotropic oscillator. This correspondence indicates that, in solving an 
 explicit model, one should always look for the simplest form
 of the solution. As far as two dimensional models are concerned,
 choice of particular solution may seem of less importance. However, 
 it becomes very important in higher dimensions where the set of equations is 
 far more complicated and finding a simple way of solving it becomes 
 essential.\\
 In this note we are going to adopt the philosophy of simplicity and
 point out its advantage in describing the three dimensional noncommutative 
 harmonic oscillator.\\
 As an introduction we are going to give a  brief review of the way 
 in which  noncommutative system can be transformed into an equivalent 
 commutative form. This approach is shown to be equivalent to the introduction
 of the  Moyal $\ast$-product \cite{nekra},\cite{presn}, \cite{gamb}, which is 
 the usual way to introduce noncommutativity.
  One starts with the set of  noncommutative coordinates $(\, x\ , p\,)$ of 
  position and momentum satisfying the following commutation relations
  
   \begin{eqnarray}
&&\left[\, x_k\ , x_j\,\right]=i\,\Theta_{kj}\label{comm1}\\
&&\left[\, p_k\ , p_j\,\right]=i\,B_{kj}\label{comm2}\\
&&\left[\, x^k\ , p_j\,\right]=i\,\delta^k{}_j\label{comm3}
\end{eqnarray}

where, ${\mathbf \Theta}$ and   ${\mathbf B}$ are matrices whose elements
measure the noncommutativity of coordinate and momenta respectively.
  We shall represent noncommutative variables as a linear combination of
  commutative coordinateses $(\,\vec \alpha\ ,\vec \beta\,)$ in a six 
  dimensional phase  space 
  
  \begin{equation}
\left(\begin{array}{c}
\vec x\\ \vec p
\end{array}\right)
=\left(\begin{array}{cc}
{\mathbf a} & {\mathbf b}\\
 {\mathbf d} &  {\mathbf c}
\end{array}\right)
\left(\begin{array}{c}
\vec \alpha \\ \vec \beta
\end{array}\right)
\end{equation}

 The $6\times 6$ transformation matrix is written in terms of 
 $3\times 3$ blocks ${\mathbf a}$, ${\mathbf b}$,  ${\mathbf c}$,
 ${\mathbf d}$. The four sub-matrices satisfy 

\begin{eqnarray}
&& {\mathbf a}\, {\mathbf b}^{\mathrm{T}} -{\mathbf b}\,{\mathbf a}^{\mathrm{T}} 
= 
\mathbf{\Theta} \label{m1}\\
&& {\mathbf c}\, {\mathbf d}^{\mathrm{T}} -{\mathbf d}\,{\mathbf c}^{\mathrm{T}} 
= 
-\mathbf{ B}\label{m2}\\
&& {\mathbf c}\, {\mathbf a}^{\mathrm{T}} -\mathbf{ b}\,{\mathbf d}^{\mathrm{T}} 
= 
\mathbf{ I}\label{m3}
\end{eqnarray}

following from the commutation relations (\ref{comm1}), (\ref{comm2}),
(\ref{comm3}).
$\mathbf{M}^{\mathrm{T}}$ denotes the transposed of a $3\times 3$ matrix
$\mathbf{M}$.\\
As a specific model we choose a three dimensional, noncommutative,  harmonic 
oscillator described by the Hamiltonian
  
\begin{equation}
H\equiv {1\over 2}\,\left[\, p_i^2 +  x_i^2\,\right]\label{h}
\end{equation}

where, we set classical frequency and mass to unity.
 One can verify that the attempt to solve the  system of equations
   (\ref{m1}),   (\ref{m2}), (\ref{m3}) in full generality
( meaning the most general form of the transformation matrices ), already in
two dimensions, led to a complicated, but still tractable, set of equations
\cite{nair},\cite{noi}.  
In three dimensions things only get considerably worse. Thus, we apply the
above mentioned philosophy of simplicity. \\
  First of all, we notice that in 
three dimensions antisymmetric matrices $\mathbf{\Theta}$ and ${\mathbf b}$
 can always be written as 

\begin{equation}
\Theta_{ab}\equiv \epsilon_{abc}\, \theta_c \ ,\qquad
B_{ab}\equiv \epsilon_{abc}\,B_c
\end{equation}

On physical grounds the isotropic solution, having spherical symmetry,
requires the equivalence of all three directions. Therefore, let us impose 
$\theta_c\equiv \theta$ and $B_c\equiv  B $, $\forall$ $c$.
Furthermore, in analogy with two dimensions, let us choose  
 matrices ${\mathbf a}$, ${\mathbf b}$, ${\mathbf c}$, ${\mathbf d}$  to be 
 
 \begin{eqnarray}
&&{\mathbf a}= a\,{\mathbf I}\ ,\qquad 
{\mathbf c}= c\,{\mathbf I}\nonumber\\
&&{\mathbf b}= b\,{\mathbf K}\ ,\qquad 
{\mathbf d}= d\,{\mathbf K}
\label{ansatz}
\end{eqnarray}

where, ${\mathbf I}$ is the identity  and ${\mathbf K}$ is an
unknown matrix. Explicit form of   ${\mathbf K}$ is found to be 

\begin{equation}
{\mathbf K}= 
\left(\, \begin{array}{ccc}
0 & 1 &0\\
0 & 0 & 1\\
1 &0 &0
\end{array}
\,\right)
\end{equation}
 
Inserting  the ansatz (\ref{ansatz}) in (\ref{m1}) and (\ref{m2}) 
leads to the solutions for the parameters as:

\begin{equation}
 b=-{\theta\over a}\label{bza}\ ,\qquad
 d={B\over c}
\end{equation}

The remaining  equation (\ref{m3}) gives

\begin{equation}
a\, c+ {B\theta\over a\, c}=1
\end{equation}

which determines the parameters $c$ and $d$ as: 

\begin{eqnarray}
&& c= {1\over 2a}\left(\,1+\sqrt{1-4\theta B}\,\right)\equiv {1\over
2a}\left(\,1+\sqrt{\kappa}\,\right)
\label{cza}\\
&& d= {a\over 2\theta}\left(\,1-\sqrt{1-4\theta B}\,\right)\equiv  
{a\over 2\theta}\left(\,1-\sqrt{\kappa}\,\right)
\label{dza}
\end{eqnarray}

The three dimensional solutions follows the same pattern as in two 
dimensions \cite{noi}.
The two dimensional isotropic representation, is  characterized by
the presence of a mixed term in the Hamiltonian, which is the reminiscence
of the noncommutativity of the system. We find that such  term
is also present in this case and is of the form

\begin{equation}
H_{mixed}= \frac{1}{2}
\left(\,-\theta\, \alpha_1\, \beta_2 + B \, \alpha_2\, \beta_1\,\right)+\dots
\label{hmix}
\end{equation}

In \cite{noi} mixed term led to the  
coupling of the noncommutative parameters to the components of the
angular momentum operator. Thus, we were able to interpret
the noncommutativity as a ``magnetic effect''.
In order to reproduce, if possible, the same interpretation
in (\ref{hmix}) one has to impose the condition 

\begin{equation}
B=\theta
\label{thetabb}
\end{equation}

which allows to rewrite the mixed term as

\begin{equation}
H_{mixed}=-\frac{1}{2} \theta_i \, L_i
\end{equation}

where, $\vec L$ is the angular momentum operator. We arrive at the isotropic
representation of the noncommutative three-$D$ harmonic oscillator 

\begin{eqnarray}
 && H= h_\alpha\, \left(\, \alpha_i\, \right)^2 + h_\beta\, \left(\, \beta_i\, 
\right)^2
 -\frac{1}{2}\vec \theta\cdot \vec L
 \label{sim}\\
 && h_\alpha\equiv {a^2\over 2}\,\left[\, 1+{1\over 4\theta^2}\,\left(\, 1 
 -\sqrt\kappa\,\right)^2\,\right]\label{h1}\\
 && h_\beta\equiv {\theta^2\over 2a^2}\,\left[\, 1+{1\over 4\theta^2}\,
 \left(\, 1 +\sqrt\kappa\,\right)^2\,\right]\label{h2}
 \end{eqnarray}

where, $\kappa\equiv 1-4\theta^2$. Hamiltonian (\ref{h1}) is invariant under
spatial rotations. This fact permits to choose a new set of coordinates with
one axis aligned with $\vec \theta$. In the rotated frame $ \alpha_i\to
{\mathbf R}_{ij }\alpha_j$, $ \beta_i\to{\mathbf R}_{ij }\beta_j$. 
The Hamiltonian (\ref{sim}) takes  a simpler looking form 

\begin{equation}
H= h_\alpha\, \left(\, \alpha_i\, \right)^2 + h_\beta\, \left(\, \beta_i\, 
\right)^2
 -\frac{\sqrt 3}{2}\, \theta \, L_\theta
\end{equation}

The explicit form of the rotation matrix is given by 

\begin{equation}
{\mathbf R}=\frac{1}{\sqrt 6}\left(\begin{array}{ccc}
\sqrt 2 & \sqrt 2 & \sqrt 2\\
-\sqrt 3 & \sqrt 3  & 0\\
-1 & -1 &  2
\end{array}\right)
\end{equation}

The spectrum of the system is

   \begin{equation}
    E_{n_+\, n_-}= \omega\left(\, n_+ + n_- +n_0  +{3\over 2 }\,\right) 
    +\left(\, n_+ - n_-\, \right)\, \frac{\sqrt 3}{2}\theta  \label{e+-}
    \end{equation}

where, $m\equiv n_+ - n_-$ is the ``magnetic'' eigenvalue of the $L_\theta$
component of the angular momentum operator, and

\begin{equation}
         \omega\equiv  2\sqrt{h_\alpha\, h_\beta}
        \label{w}
        \end{equation}

$\omega$ being expressed in terms of units of  classical frequency.
The explicit solution give the frequency of the harmonic oscillator equal
to the classical frequency. The noncommutative effects are  pure
magnetic field effects in (\ref{sim}). The results are identical to two 
dimensional case for the special choice $\theta=B$. In three dimensions, 
however, this choice is imposed
by the form of the mixed Hamiltonian and is the only possible solution which
gives isotropy of the Hamiltonian. One can re-write the spectrum in the
following way
 
\begin{equation}
    E_{n_+\, n_-}= \omega_+\left(\, n_+ + {1\over 2 }\,\right) +
    \omega_-\left(\, n_- + {1\over 2 }\,\right)+\omega\left(\, n_0 + 
    {1\over 2}\,\right)
    \end{equation}

provided the following identifications are in order

\begin{eqnarray}
&&\sqrt 3 \, \theta= \left(\, \omega_+ - \omega_-\,\right)\\
&& \omega=\frac{1}{2}\left(\, \omega_+ +\omega_-\,\right)
\end{eqnarray}

The above spectrum is the one of three, one-dimensional, anisotropic 
oscillators. Thus, $3D$ 
noncommutative harmonic oscillator has both isotropic and anisotropic
commutative representations. In order to prove the existence of solutions
for the transformation matrices of the anisotropic representation, without
explicitly solving complex set of equations (\ref{m1}), (\ref{m2}), (\ref{m3}),
one can  proceed in the following way. Let us first find the relation among the 
commutative coordinates of the two different representations by defining 

\begin{eqnarray}
&& Q_1= A_1\, \alpha_1 -A_2\, \beta_2\ ,\qquad 
 Q_2= -A_1\, \alpha_2 +A_2\, \beta_1\ ,\qquad
 Q_3= C \alpha_3\\ 
&& P_1= A_1\, \alpha_2 +A_2\, \beta_1\ ,\qquad
 P_2= -A_1 \, \alpha_1 -A_2\, \beta_2\ ,\qquad
 P_3= D \,\beta_3\label{pq}
\end{eqnarray}

The parameters in (\ref{pq}) are determined by the requirement that 
the above redefinitions turn the Hamiltonian (\ref{sim}) into its anisotropic
form  

\begin{equation}
H=\frac{1}{2}\omega_+\left( Q_1^2 + P_1^2\right) + 
\frac{1}{2} \omega_-\left( Q_2^2 + P_2^2\right)
+\frac{1}{2}\omega\left( Q_3^2 + P_3^2\right) 
\end{equation}

which gives the solutions

\begin{eqnarray}
&& A_1= \sqrt{\frac{h_\alpha}{\omega}}\ ,\qquad
   A_2= \sqrt{ \frac{h_\beta}{\omega}}\nonumber\\
&& C=\sqrt{2 }A_1\ ,\qquad
 D=\sqrt{2} A_2
\label{abcd}
\end{eqnarray}

The relation between the anisotropic coordinates $\left(\, \vec Q\ , \vec P
\,\right)$, written as a ``column matrix'' $ \mathbf{Q}$, and isotropic
ones $\left(\, \vec\alpha\ , \vec\beta\,\right)$ can be written in matrix form 
as

\begin{equation}
\left(\begin{array}{c}
\vec\alpha\\ \vec\beta
\end{array}\right)
=\left(\begin{array}{cc}
A_2\, {\mathbf L}_1 & A_2\, {\mathbf L}_2\\
-A_1\, {\mathbf L}_2 &  A_1\, {\mathbf L}_1
\end{array}\right)
\left(\begin{array}{c}
\vec Q \\\vec P
\end{array}\right)
\label{abPQ}
\end{equation}

The above equation is written in the block form with $3\times 3$ matrices
${\mathbf L}_1$, ${\mathbf L}_2$ given by

\begin{equation}
 {\mathbf L}_1=\left(\begin{array}{ccc}
1 & 0 & 0\\
0 & -1 & 0\\
0 & 0 & \sqrt 2
\end{array}\right) \ ,\qquad
 {\mathbf L}_2=\left(\begin{array}{ccc}
0 & -1 & 0\\
1 & 0 & 0\\
0 & 0 & 0
\end{array}\right)
\end{equation}

The relation between the noncommutative coordinates ( after
rotation ) and  the isotropic set of solutions can be written in a block form 

\begin{equation}
\left(\begin{array}{c}
\vec x\\ \vec p
\end{array}\right)
=
\left(\begin{array}{cc}
a\,{\mathbf R}^T & b\,{\mathbf K}{\mathbf R}^T \\
d\,{\mathbf K}{\mathbf R}^T & c \,{\mathbf R}^T
\end{array}\right)\left(\begin{array}{c}
\vec \alpha \\ \vec \beta
\end{array}\right)
\label{xpab}
\end{equation}

On the other hand, the anisotropic transformation matrices  relate the 
noncommutative coordinates to $\left(\, \vec Q\ , \vec P\,\right)$ as

\begin{equation}
\left(\begin{array}{c}
\vec x\\ \vec p
\end{array}\right)
=\left(\begin{array}{cc}
\widetilde{\mathbf a} & \widetilde{\mathbf b}\\
\widetilde {\mathbf d} &  \widetilde{\mathbf c}
\end{array}\right)
\left(\begin{array}{c}
\vec Q \\\vec P
\end{array}\right)
\label{xpQP}
\end{equation}

Comparing (\ref{xpQP}) to  (\ref{xpab}) with the help of (\ref{abPQ}) one
 obtains solutions

\begin{eqnarray}
&&\widetilde{\mathbf a}=  a\, A_2\,\widetilde {\mathbf L}_1- b\, A_1\,
 \widetilde{\mathbf L}_4\ ,\qquad
 \widetilde{\mathbf b}=  a\, A_2 \,\widetilde {\mathbf L}_2+ b\,  A_1 \,
 \widetilde{\mathbf L}_3\\
&&\widetilde{\mathbf c}=  d\, A_2\, \widetilde {\mathbf L}_4+ c\,  A_1\,
 \widetilde{\mathbf L}_1\ , \qquad
\widetilde{\mathbf d}=  d\, A_2\, \widetilde {\mathbf L}_3- c\,  A_1\,
 \widetilde{\mathbf L}_2
 \end{eqnarray}
 
 where the $3\times 3$ matrices $\widetilde{\mathbf L}_i$, $i=1\ , 2\ , 3\ ,4$,
 are found to be
 \begin{eqnarray}
 &&\widetilde{\mathbf L}_1=\frac{1}{\sqrt 6}\left(\begin{array}{ccc}
\sqrt 2 & \sqrt 3 & -\sqrt 2\\
\sqrt 2 & -\sqrt 3  & -\sqrt 2 \\
 \sqrt 2 & 0 &  2\sqrt 2
\end{array}\right)\ ,\qquad
\widetilde{\mathbf L}_2=\frac{1}{\sqrt 6}\left(\begin{array}{ccc}
-\sqrt 3 & -\sqrt 2 & 0\\
\sqrt 3 & -\sqrt 2  & 0 \\
 0 &   -\sqrt 2  & 0
\end{array}\right)\\
&&\widetilde{\mathbf L}_3=\frac{1}{\sqrt 6}\left(\begin{array}{ccc}
\sqrt 2 & -\sqrt 3 & -\sqrt 2\\
\sqrt 2 & 0  & 2\sqrt 2 \\
 \sqrt 2 &  \sqrt 3  &  -\sqrt 2
\end{array}\right)\ ,\qquad
\widetilde{\mathbf L}_4=\frac{1}{\sqrt 6}\left(\begin{array}{ccc}
\sqrt 3 & -\sqrt 2 & 0\\
0 & -\sqrt 2  & 0\\
 -\sqrt 3 &  -\sqrt 2  &  0
\end{array}\right)
\end{eqnarray}

Exploiting the explicit solutions (\ref{h1}), (\ref{h2}) and (\ref{abcd}),
 one can re-write the anisotropic set of solutions in terms of isotropic ones

\begin{eqnarray}
&&\widetilde{\mathbf a}=  \sqrt{\frac{ac}{2}}\,\left(\,
\widetilde{\mathbf L}_1+\frac{\theta}{ac}\,
\widetilde{\mathbf L}_4\,\right)\ ,\qquad
\widetilde{\mathbf b}=  \sqrt{\frac{ac}{2}}\left(\, \widetilde{\mathbf L}_2-
\frac{\theta}{ac}\,\widetilde{\mathbf L}_3\,\right)\nonumber\\
&&\widetilde{\mathbf a}= \widetilde{\mathbf c}\ ,\qquad
\widetilde{\mathbf d}= -\widetilde{\mathbf b} \label{anis}
 \end{eqnarray}
 
 One can verify that the above set of solutions satisfies basic requirements
 (\ref{m1}), (\ref{m2}), (\ref{m3}).  
 As already advocated, a comparison of the isotropic solutions to the
 anisotropic ones confirms the validity of the ``philosophy of simplicity''
 approach. \\
 One may wonder if there are other solutions leading to the same isotropic
 representation. Let us assume that the transformation matrices are
 of the same form as before, but with different matrix elements.
 For example, the matrices are $\textbf{a}$ and  $\textbf{c}$
 are $\textbf{a}_{ij}=a^{(i)}\,\delta_{ij}$ ( no 
summation over the  $i$ index ),while  matrices $\textbf{b}$ and  $\textbf{d}$ 
are given by

\begin{equation}
{\mathbf b}= 
\left(\, \begin{array}{ccc}
0 & b_{12} &0\\
0 & 0 &  b_{23}\\
b_{31} &0 &0
\end{array}
\,\right)
\ ,\qquad
{\mathbf d}= 
\left(\, \begin{array}{ccc}
0 & d_{12} &0\\
0 & 0 &  d_{23}\\
d_{31} &0 &0
\end{array}
\,\right)
\end{equation}

Without going into details, the Hamiltonian following from the above solution is
the generalization of (\ref{sim}) with different coefficients $h_i$, $i=1\ ,
\dots\, 6$ multiplying canonical coordinates. The isotropy of the system
requires the equivalence of those coefficients for the coordinates $\alpha$
and $\beta$ respectively. This requirement inevitably leads to the condition
 (\ref{thetabb}). Thus, we conclude that there are no other 
isotropic solutions different from those described in this paper.
We have thus shown that the three-$D$ noncommmutative harmonic oscillator can 
 be represented as an isotropic oscillator coupled to an external
 magnetic field, generated by  space non-commutativity.  This representation
 is based on a very simple set of transformation matrices relating
 noncommutative to canonical coordinates. Alternative representation is also
 possible in terms of three $1D$ anisotropic harmonic oscillators. The second
 set of solutions is far more complicated and difficult to obtain solving
 (\ref{m1}), (\ref{m2}), (\ref{m3}). Nevertheless, we have described an
 indirect way of finding these solutions. Their explicit form was needed
 to support the philosophy of simplicity approach described in this paper.

\end{document}